\documentclass[12pt,aps,showpacs,preprintnumbers,onecolumn,superscriptaddress,floatfix,nofootinbib]{article}

\usepackage{amssymb}
\usepackage{amsmath}
\usepackage{comment}
\usepackage[T1]{fontenc}
\usepackage{graphicx}
\usepackage{rotating}
\usepackage{hyperref}
\usepackage{soul,color}
\usepackage{booktabs}
\usepackage{longtable}
\usepackage{authblk}

\providecommand{\keywords}[1]{\textbf{\textit{Keywords---}} #1}

\begin{document}
	
	
	
	
	\title{Statistical Mechanics of the Sub-Optimal Transport} 
	\author{Riccardo Piombo$^{1}$,  
		Lorenzo Buffa$^{1}$, 
		Dario Mazzilli$^{1}$, 
		Aurelio Patelli$^{1}$
		\\  
		$^{1}$Enrico Fermi Research Center (CREF), Rome (Italy)}
	
	\maketitle
	
	\begin{abstract}
		Statistical mechanics is a powerful framework for analyzing optimization yielding analytical results for matching, optimal transport, and other combinatorial problems. 
		However, these methods typically target the zero-temperature limit, where systems collapse onto optimal configurations, a.k.a. the ground states. 
		Real-world systems often occupy intermediate regimes where entropy and cost minimization genuinely compete, producing configurations that are structured yet sub-optimal. 
		The Sub-Optimal Transport (SOT) model captures this competition through an ensemble of weighted bipartite graphs: a coupling parameter interpolates between entropy-dominated dense configurations and cost-dominated sparse structures. 
		This crossover has been observed numerically but lacked analytical understanding. 
		Here we develop a mean-field theory that characterizes this transition. 
		We show that local fluctuations in Lagrange multipliers become sub-extensive in the thermodynamic limit, reducing the full model with strength constraints to an effective single-constraint problem admitting an exact solution in some intermediate regime. 
		The resulting free energy is analytic in the coupling parameter, confirming a smooth crossover rather than a phase transition. 
		We derive closed-form expressions for thermodynamic observables and weight distributions, validated against numerical simulations. 
		These results establish the first analytical description of the SOT model, extending statistical mechanics methods beyond the zero-temperature regime.
	\end{abstract}

		
		

	
	
\section{Introduction}
\label{sec:intro}

Statistical mechanics provides a powerful framework for analyzing optimization problems, enabling both efficient algorithms and analytical characterizations of optimal solutions. The connection arises naturally: any optimization problem can be cast as energy minimization, where the cost (objective) function serves as a Hamiltonian. This correspondence developed into a rich body of analytical techniques (including the replica method, cavity equations, and mean-field theories) originally designed for disordered systems such as spin glasses \cite{MezardParisiVirasoro1987, MartinMonassonZecchina2001}.

These methods have also been successfully applied to a broad class of combinatorial optimization problems defined on graphs. 
Among the most studied are matching problems, where the goal is to pair elements across two sets at minimum cost. 
Mézard and Parisi applied the replica method to compute the expected cost of optimal matchings in both the low-temperature limit and the thermodynamic limit ~\cite{Mezard1985, Mzard1988}, while Aldous found a rigorous derivation in the same regime~\cite{aldous2001zeta}. 
Related problems include stable marriage and the roommate problem, where preferences impose additional stability constraints \cite{Fenoaltea2021}, and the traveling salesman problem, which seeks an optimal cycle through a set of cities \cite{MezardParisi1986TSP, Vannimenus1984}. 
More recently, \cite{Baybusinov2025} reformulated matching within a grand-canonical framework, recovering the classical Mézard-Parisi results while sharing some mathematical tools with the model described in this work.

Optimal transport (OT) extends this family of problems by replacing binary assignments with continuous mass distributions subject to marginal constraints~\cite{OT_general2,PeyreCuturi2019}. 
OT seeks a coupling that specifies how much mass flows along each edge while satisfying prescribed row and column sums. 
\cite{Koehl2019PRE, Koehl2019, koehl2024} developed a statistical-physics approach to discrete OT using saddle-point methods, deriving a formulation that yields algorithms competitive with entropy-regularized methods such as \cite{Cuturi2013}. 

A common feature unites these analyses: they target the zero-temperature limit, where the system collapses onto a single optimal configuration amenable to exact treatment.
Real-world systems, however, rarely achieve perfect optimization. 
Configurations that are sub-optimal yet structured, neither fully random nor fully optimized, emerges frequently. 
The analytical machinery developed for zero-temperature problems does not directly apply to this intermediate regime, because taking this limit before the thermodynamic limit eliminates precisely the sub-optimality of interest: the thermodynamic limit and the zero-temperature limit do not commute.
A statistical mechanics framework capable of describing systems at finite temperature, where entropy and cost minimization genuinely compete, remains a gap in the literature.

The Sub-Optimal Transport (SOT) model, introduced in \cite{buffa2025}, provides a natural setting to address this challenge. 
The model defines an ensemble of weighted bipartite graphs in which a tuning parameter $\beta$ interpolates continuously between entropy-dominated (dense, random) and cost-dominated (sparse, structured) configurations. 
Numerical simulations reveal a well-defined crossover between these regimes, yet no analytical theory has captured this transition within a statistical mechanics framework. 
Providing such a theory would clarify how large-scale structural organization emerges from the competition between randomness and optimization.

In this work, we derive a mean-field theory for the SOT model that analytically characterizes the dense-to-sparse crossover. 
Starting from the partition function with node-wise strength constraints, we show that local fluctuations in the Lagrange multipliers become sub-extensive in the thermodynamic limit, reducing the full model to an effective single-constraint problem admitting exact solution. 
The resulting free energy is analytic in $\beta$, confirming that the transition is a smooth crossover rather than a true phase transition. 
We validate the theory by comparing predicted thermodynamic observables (mean energy, susceptibility, and weight distributions) against numerical solutions across a range of system sizes and cost distributions. 
Our results establish the first analytical description of sub-optimal transport and demonstrate how mean-field methods can be extended beyond the zero-temperature regime.

\section{Sub-OT model and the relation between stat-mech and maximum entropy}
\label{sec:model}

The SOT model defines a probability distribution over mass configurations on a bipartite network by maximizing Shannon entropy subject to marginal constraints. The framework imposes mass conservation through strength vectors $ (s, \sigma)$, ensuring that total mass is conserved at every node:
\begin{equation}
	\label{eq:marginals}
	\sum_\alpha W_{i\alpha} = s_i \qquad \sum_i W_{i\alpha} = \sigma_\alpha
\end{equation}
These constraints define an ensemble of random weighted bipartite graphs, where each $W_{i\alpha}$ represents mass on the link connecting node $i$ in one layer to node $\alpha$ in the other. The cost matrix $C_{i\alpha}$ enters not as a constraint but through Boltzmann-like weighting, yielding the following probability distribution over mass configurations:
\begin{equation}
	\label{eq:SOT_probability}
	P_{SOT} = \frac{1}{Z_{SOT}} e^{-\sum_{i,\alpha} W_{i\alpha}(\beta \,C_{i\alpha} +\lambda_i + \mu_\alpha)}
\end{equation}
where $Z_{SOT}$ is the partition function ensuring normalization and $\{\lambda_i,\mu_\alpha\}$ are the Lagrange multipliers relative to constraints in Eq.(\ref{eq:marginals}).

Configurations with lower transport cost $\sum_{i,\alpha} C_{i\alpha} W_{i\alpha}$ receive higher probability, with the coupling parameter $\beta$ controlling the strength of this bias. 

This formulation extends the maximum-entropy approach for random graphs~\cite{cimini2019} by introducing the cost matrix as an external field that competes with entropy maximization. At small $\beta$, entropy dominates and mass distributes nearly uniformly across edges; at large $\beta$, the distribution concentrates on configurations that minimize transport cost, yielding sparse, structured networks. The $\beta$ values in between capture intermediate stages of optimization.

Numerical studies reveal that the SOT model undergoes a non-critical transition (crossover) between two distinct regimes as $\beta$ varies. In the \emph{dense phase} at small $\beta$, weights distribute approximately homogeneously across all links except where constraints force specific patterns. In the \emph{sparse phase} at large $\beta$, weights concentrate onto a limited subset of links forming a \emph{spanning tree} structure whose size scales linearly with system size. This spanning tree represents a low-entropy, highly ordered state emerging from optimization.

The \emph{dense-to-sparse} transition can be quantified through an order parameter: the fraction of total mass contained within the maximum spanning tree as $\beta$ varies. This fraction vanishes in the dense phase (approaching zero for large systems) and approaches unity in the sparse limit.
Because the spanning tree configuration emerges in the large-$\beta$ limit and the coupling parameter effectively varies on an exponential scale, $\ln \beta$ provides the natural variable for mapping the transition. The transition exhibits well-defined but non-critical character: the order parameter's first derivative reaches a pronounced maximum rather than diverging, indicating a smooth crossover rather than a true phase transition. Moreover, the transition point remains stable across different cost distributions, demonstrating robustness.

The SOT framework can be formulated equivalently using statistical mechanics. This representation interprets the cost term as an external field acting on the mass field $W$, while strength constraints enter as Dirac delta functions. By introducing Fourier variables $\{\lambda_i, \mu_\alpha\}$ conjugate to these constraints, the partition function takes the form:
\begin{eqnarray}
	Z\left(\beta,\{s_i\},\{\sigma_\alpha\}\right)&=&\int \prod_{i,\alpha}  dW_{i\alpha}\,\,\,\,e^{-\beta \sum_{i,\alpha}W_{i\alpha}C_{i\alpha}} \times\nonumber\\
	&\,&\prod_{i} \delta\!\left(\sum_\alpha W_{i\alpha} - s_i\right)\,\,\,\prod_{\alpha} \delta\!\left(\sum_i W_{i\alpha} - \sigma_\alpha\right)\nonumber\\
	&\approx&\int \mathcal{D}\lambda\, \mathcal{D}\mu\,\mathcal{D}W\,\exp\left\{ \imath \sum_i\lambda_is_i +  \imath \sum_\alpha\mu_\alpha\sigma_\alpha  \right.\nonumber\\
	&\,&\left. - \sum_{i,\alpha}W_{i\alpha}(\beta C_{i\alpha} + \imath \lambda_i +\imath \mu_\alpha) \right\}
	\label{eq:zsot}
\end{eqnarray}
where $\mathcal{D}\lambda = \prod_i d\lambda_i$ and $\mathcal{D}\mu = \prod_\alpha d\mu_\alpha$ are the measures of the Fourier variables associated to Dirac deltas and $\mathcal{D}W = \prod_{i,\alpha}  dW_{i\alpha}$. $Z$ quantifies the occupation of accessible phase space given the model parameters. This formulation imposes constraints as \emph{hard} requirements: every configuration $W$ must exactly satisfy the marginal conditions in Eq.\eqref{eq:marginals}.

The information-based approach relaxes hard constraints into \emph{soft} constraints enforced only on average. 
Instead of requiring exact constraint satisfaction for every configuration, this approach maximizes likelihood subject to constraints holding in expectation. This distinction parallels the difference between microcanonical and canonical ensembles in statistical mechanics: the microcanonical ensemble fixes energy exactly (hard constraint), while the canonical ensemble fixes only average energy (soft constraint). Moving from hard to soft constraints is analogous to moving from the microcanonical to the canonical ensemble in statistical mechanics, as discussed in~\cite{pathria1996statisticalmechanics,zhang2022strong,squartini2015breaking}.

Having established this parallel between microcanonical/canonical ensembles and hard/soft constraints, we can now explore the statistical mechanics perspective of the model. 
The two frameworks can be compared in terms of their integration procedures over the system's degrees of freedom. Defining the free energy density as:
\begin{equation}
	\label{eq:Free-energy-density}
	\mathcal{F}(\lambda,\mu,W\vert\beta,s,\sigma)= \sum_{i,\alpha}W_{i\alpha}(\beta C_{i\alpha} + \imath \lambda_i +\imath \mu_\alpha)-\imath \sum_i\lambda_is_i - \imath \sum_\alpha\mu_\alpha\sigma_\alpha
\end{equation}
provides a natural point of comparison emerging from the canonical structure of Eq.(\ref{eq:zsot}).
The statistical mechanics approach evaluates the partition function $Z$ by integrating over all variables ($W$, $\lambda$, and $\mu$) thereby accounting for all configurations consistent with constraints.
In contrast, the information-based approach integrates only over $W$ while fixing $\lambda$ and $\mu$ at their most probable values (the maximum of the log-likelihood). This constitutes an approximation that converges to the full $Z$ only when $\mathcal{F}$ in Eq.(\ref{eq:Free-energy-density}) exhibits a dominant peak around the most probable configuration.

Equivalence between the two approaches requires specific conditions that are not universally satisfied. 
In the absence of a diverging parameter, such as the number of particles in conventional statistical mechanics~\cite{pathria1996statisticalmechanics}, the peak approximation may fail~\footnote{The maximum of a probability distribution provide the mode, which is not always a good estimate of the mean. 
	For asymmetric distributions, only when the variance vanishes do mode and mean converge to the same value.}, and the information-based description may deviate from the statistical mechanics treatment. This distinction becomes crucial when comparing theoretical predictions with numerical results from either framework.

\section{Mean Field analysis in the case of uniform cost distribution}
\label{sec:mean_field}

In this section we evaluate the partition function within the mean-field approximation, assuming that the cost matrix entries are drawn from a uniform distribution on the interval $[0, 1]$. We begin with a simplified case where the complete set of node-wise constraints is replaced by a single global mass constraint. Subsequently, we extend the analysis to the full model with the complete set of local constraints $(s,\sigma)$.

\subsection{Total mass conservation}
\label{sec:one_constraint}
We first consider a simplified case where only a global constraint on total mass is imposed. 
Defining the total mass as $\bar{W} = \sum_\gamma W_\gamma$, where $\gamma$ indexes all node pairs $(i,\alpha)$, the partition function becomes:
\begin{subequations}
	\begin{eqnarray}
		Z&=&\int \mathcal{D}W\,e^{-\beta\sum_\gamma W_\gamma C_\gamma}\delta\left(\bar{W} - \sum_\gamma W_\gamma\right) \label{eq:Z_total_mass}\\
		&=& \frac{1}{2\pi}\int d\lambda \,e^{\imath\lambda \bar{W}}\prod_\gamma \int dW e^{- W(\beta C_\gamma + \imath \lambda)}\nonumber\\
		&=& \frac{1}{2\pi}\int d\lambda\, e^{\imath\lambda \bar{W}}\prod_\gamma \frac{1}{\beta C_\gamma + \imath \lambda}\label{eq:er_constraint}\\
		&=& \frac{1}{\beta^{N_\gamma-1}} \sum_\gamma (-1)^\gamma e^{-\beta\bar{W}C_\gamma} \prod_{\gamma'\neq \gamma} \frac{1}{\vert C_{\gamma'} - C_\gamma\vert} \label{eq:4c}
	\end{eqnarray}
\end{subequations}
The last equation follows from countour integration: the integrand has simple poles at $\lambda = -\imath \beta C_\gamma$. 

We evaluate the \emph{configurational average} of the partition function, given a prescribed distribution of the \emph{disorder} $C\leftarrow \rho(c)$.
By considering the $C_\gamma$ in increasing order of the
index $\gamma$, we may assume without loss of generality that $C_\gamma<C_{\gamma+1}$, for $\gamma\in[1,\cdots,N_\gamma]$.
When the costs are independently drawn from a uniform distribution, the ordered variables become self-averaging in the large-$N_\gamma$ limit, so that fluctuations around their mean are suppressed and the value of the configurational average can be exactly calculated.
Under this assumption, we may approximate each ordered cost as its expected value, yielding $C_\gamma\sim\mathbb{E}[C_\gamma]=\frac{\gamma}{N_\gamma+1}\,$, which constitutes a mean-field approximation~\footnote{This expression corresponds to the expected value of the $\gamma$-th order statistic among $N_\gamma$ independent samples drawn from a uniform distribution.}.
It follows that, in the large $N_\gamma$ limit, the partition function can be expressed as (the complete calculation is reported in \ref{app:derivation_Z_mean_field}):
\begin{subequations}
	\begin{eqnarray}
		Z&\simeq& \frac{1}{\beta^{N_\gamma}} \sum_\gamma (-1)^\gamma e^{-\beta\bar{\omega}\gamma} \frac{N_\gamma!}{\gamma!(N_\gamma-\gamma)!} = \frac{1}{\beta^{N_\gamma}}\left(1-e^{-\beta\bar{\omega}}\right)^{N_\gamma} \nonumber\\
		&=& e^{N_\gamma\mathrm{F}(\beta,\bar{\omega})} \label{eq:Z_mean_field}
	\end{eqnarray}
\end{subequations}
Here $\bar{\omega}= \frac{\bar{W}}{N_\gamma}$ is the \emph{average mass per edge} and the free energy is:
\begin{equation}
	\mathrm{F}(\beta,\bar{\omega}) =\left[\ln\left(1-e^{-\beta\bar{\omega}}\right) -\ln(\beta)  \right]
	\label{eq:free_energy_global}
\end{equation}

This exact solution reveals two fundamental features that characterize the Sub-Optimal Transport transition.

First, the logarithmic dependence emerges directly from the solution structure in Eq.(\ref{eq:free_energy_global}), whereas in the original numerical SOT formulation it was inferred only through scaling arguments. This scaling reflects the system's exponential sensitivity to variations in the coupling parameter. 

Second, the free energy $\mathrm{F}$ remains analytic and smooth at all orders of $\beta$. The system therefore does not undergo a true thermodynamic phase transition but rather exhibits a continuous crossover between regimes, consistent with the absence of any singularities in $\mathrm{F}$.

The canonical ensemble, where the hard constraint on total mass is relaxed to a soft constraint, can be derived from Eq.\eqref{eq:er_constraint}. 
In that equation we introduced the conjugate variable $\lambda$ via Fourier transform and still contains the term $\exp({i\lambda\bar{W}})$ that enforces the constraint exactly. 
The key observation is that Eq.\eqref{eq:er_constraint} can be rewritten by collecting terms in the exponent:
\begin{equation}
	\label{eq:7}
	Z = \frac{1}{2\pi} \int d\lambda \exp\left\{i\lambda\bar{W} - \sum_\gamma \ln(\beta c_\gamma + i\lambda)\right\}
\end{equation}
where the second term originates from integrating out the weights.

In the thermodynamic limit $N_\gamma \to \infty$, we can approximate the sum over individual edge costs by an integral over the cost distribution, thereby accessing a mean-field description. Defining the empirical distribution as $\rho(c)=1/{N_\gamma}\sum_\gamma \delta(c-c_\gamma)$ we can rewrite the sum over $\gamma$ as an average:
\begin{equation}
	\label{eq:glivenko_cantelli}
	\sum_\gamma\ln\left( \beta c_\gamma + \imath\lambda \right) = N_\gamma \int \ln\left( \beta c + \imath\lambda \right) \rho(c) dc  
\end{equation}
As the number of samples grows ($N_\gamma \gg 1$), the empirical distribution converges to the theoretical distribution from which costs are drawn. This convergence occurs for uncorrelated costs allowing us to replace the empirical distribution with the theoretical one and to evaluate Eq.\eqref{eq:glivenko_cantelli} analytically:
\begin{equation}
	\int_0^1 \ln(\beta c + i\lambda) \, dc \simeq \frac{N_\gamma}{\beta}\left[(\beta+\imath\lambda)\ln(\beta+\imath\lambda) - \imath\lambda \ln \imath\lambda -\beta\right] \label{eq:mean_entropy}
\end{equation}
Thus, the partition function becomes:
\begin{equation}
	Z\approx \frac{1}{2\pi}\int d\lambda \exp\left\{ \frac{N_\gamma}{\beta}\left[\imath\beta\bar\omega\lambda + \beta - (\beta+\imath\lambda)\ln(\beta+\imath\lambda) + \imath\lambda \ln \imath\lambda\right]\right\}
\end{equation}
For $N_\gamma \gg 1$, the integrand becomes sharply peaked around its maximum. 
The integral is therefore dominated by the value of $\lambda$ that maximizes the exponent, i.e. the saddle point $\lambda^*$. 
Taking the derivative of the argument of the exponential with respect to $\lambda$ and setting it to zero yields:
\begin{equation}
	0=\imath \beta\bar\omega + \imath \ln \imath\lambda^*  - \imath\ln(\beta + \imath\lambda^*) \quad\Rightarrow \quad \imath\lambda^* = \frac{\beta}{e^{\beta\bar\omega}-1}
	\label{eq:lagramge_single_constraint}
\end{equation}
The Lagrange multiplier $\lambda$ becomes a physical field whose value controls the average mass. 
For finite $N_\gamma$, $\lambda$ fluctuates around $\lambda^*$, but as $N_\gamma \to \infty$, fluctuations are suppressed and $\lambda$ concentrates at $\lambda^*$. 

Evaluating the partition function at this saddle point yields:
\begin{eqnarray}
	Z&=& \exp\left\{ N_\gamma\left[\frac{\beta\bar\omega}{e^{\beta\bar\omega}-1} - \frac{e^{\beta\bar\omega}}{e^{\beta\bar\omega}-1}\ln(\beta\frac{e^{\beta\bar\omega}}{e^{\beta\bar\omega}-1}) + \frac{1}{e^{\beta\bar\omega}-1}\ln(\frac{\beta}{e^{\beta\bar\omega}-1})\right]\right\}\nonumber\\
	&=&\exp\left\{ N_\gamma\left[\ln\left(1-e^{\beta\bar\omega}\right)-\ln(\beta)\right]\right\} = e^{N_\gamma \mathrm{F}(\beta,\bar\omega)} \label{eq:ens_eq}
\end{eqnarray}
The result in Eq.(\ref{eq:ens_eq}) demonstrates that canonical and microcanonical versions of the SOT model become equivalent in the thermodynamic limit (\cite{touchette2009large}). 
The equivalence arises because the system size acts as a large controlling parameter, playing the same role as particles number in conventional statistical mechanics. 
Although this dependence is not immediately apparent in the full model formulation, it ensures convergence of the two ensemble descriptions in the asymptotic limit for the case of global constraints. 

\paragraph{Thermodynamic observables characterize the dense-to-sparse crossover}
\begin{figure}[!t]
	\centering
	\includegraphics[width=1\linewidth]{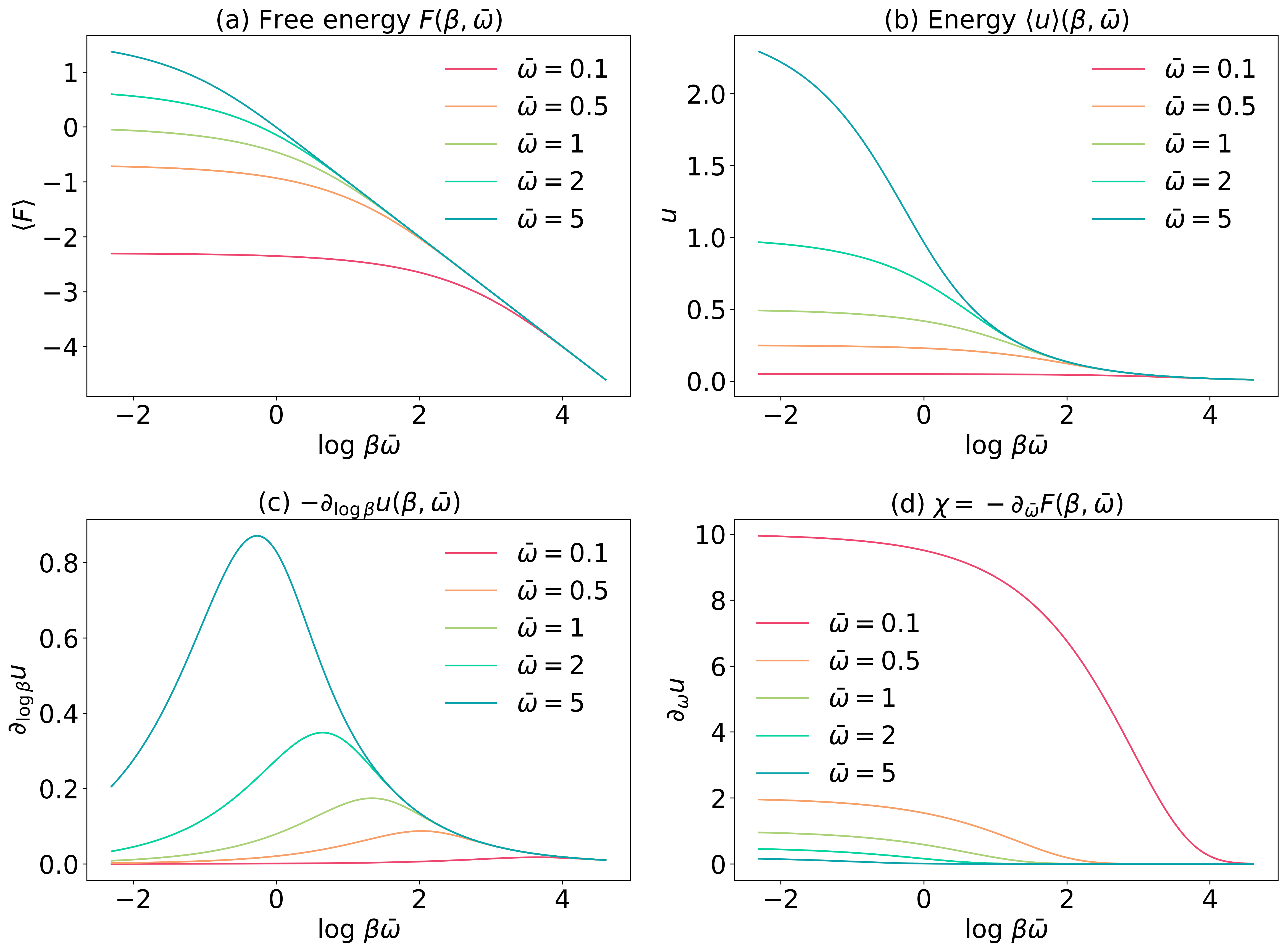}
	\caption{\textbf{Thermodynamic quantities for the single-constraint ensemble} Panel (a) Free energy of Eq.\eqref{eq:free_energy_global} vs $\log\beta\bar\omega$ for different mean mass values $\bar\omega$. Panel (b) Mean energy in Eq.\eqref{eq:energy} showing crossover between high-noise ($u = \bar \omega/2$) and low-noise ($u\to0$) regimes for different mean mass values $\bar\omega$.
		Panel (c) Energy susceptibility vs $\log\beta\bar\omega$.
		Panel (d) Mass susceptibility in Eq.(\ref{eq:mass_susceptibility}) showing system response to $\bar \omega$ variations.}
	\label{fig:theory}
\end{figure}

The free energy as a function of $\beta$ reveals how the system transitions between regimes. 
Panel (a) of Figure~\ref{fig:theory} shows the free energy from Eq.\eqref{eq:free_energy_global} as a function of $\beta$ for different constraint values $\bar \omega$. 
At large $\beta$, all curves converge to the same slope $\partial F / \partial \ln \beta \to -1$, meaning that $F\sim -\ln\beta$ becomes increasingly negative as $\beta$ grows. 
In this regime, the constraint $\bar{\omega}$ merely rescales the effective coupling but does not affect the asymptotic behavior. 
Since $Z=\exp(N_\gamma F)$, the divergence $F \to -\infty$ implies that the partition function $Z \to 0$ in a way that concentrates the probability distribution onto a single optimal configuration: the extremal solution that minimizes the cost functional as $\beta \to \infty$.

The average energy quantifies the crossover between high-noise and low-noise regimes.
The thermodynamic potential conjugate to the coupling parameter $\beta$ is the average energy $u$:
\begin{equation}
	u := \langle c W \rangle = -\frac{1}{N_\gamma} \frac{\partial F}{\partial \beta} = \frac{1}{\beta} - \bar\omega \frac{e^{-\beta\bar\omega}}{1-e^{-\beta\bar\omega}}
	\label{eq:energy}
\end{equation}
This observable, shown in panel (b) of Figure~\ref{fig:theory}, shows two distinct regimes:
In the small $\beta$ (high-noise) regime, entropy dominates and costs decorrelate from weights. 
The system approaches a uniform distribution of weights, where each edge carries approximately $\bar{\omega} \sim W_{i\alpha}/N_\gamma$ independently of its cost $C_{i\alpha}$. 
Since costs are drawn uniformly from $[0,1]$ and are independent, the average energy factorizes as $\langle cW \rangle \to \langle c \rangle \,\bar{\omega} = \bar{\omega}/2$ where $\langle c \rangle = 1/2$ is the mean of the cost distribution. 
As $\beta$ increases mass migrates from high-cost edges $(C_{i\alpha} \sim 1)$ to low-cost edges $(C_{i\alpha} \to 0)$. The average energy vanishes asymptotically, $u \to 0$, because the probability distribution allocates mass exclusively to the subset of edges with minimal cost.

In the thermodynamic limit, the energy itself remains finite and does not present discontinuities. 
Its derivative with respect to the control parameter, however, develops a pronounced peak, as shown in panel (c) of Figure~\ref{fig:theory}. 
This behavior is consistent with the absence of a genuine phase transition, as also discussed numerically in~\cite{buffa2025}. 
At the same time, because there is no true divergence, different observables can pinpoint the dense-to-sparse crossover at slightly different locations: each “order parameter” effectively defines the crossover as the value of $\beta$ where it changes most rapidly, and these points need not coincide. 
For instance, for $\bar\omega=1$ the energy derivative signal a peak at $\log \beta \simeq 1.341$, whereas the Maximum Spanning Tree mass share reported in~\cite{buffa2025} indicates a crossover at approximately $\log \beta \simeq 3.1$-$3.2$.

The variation of free energy with respect to $\bar\omega$ is directly related to the system’s response to changes in the total mass, and therefore to the emergence of structural organization in weight allocation:
\begin{equation}
	\chi(\beta,\bar\omega) = -\frac{1}{N_\gamma} \frac{\partial F}{\partial {\bar{\omega}}} = \beta \frac{e^{-\beta\bar\omega}}{1-e^{-\beta\bar\omega}} = \imath\lambda^*
	\label{eq:mass_susceptibility}
\end{equation}
Panel (d) of Figure~\ref{fig:theory} shows that for very large $\beta$, the response converges to zero. This behavior is consistent with the observation that the probability distribution in network space rigidly condenses toward a single structure, causing fluctuations to vanish. This susceptibility to the total mass variation equals the Lagrange multiplier under the Saddle Point Approximation in Eq.~\eqref{eq:lagramge_single_constraint}.
It is analogous to the chemical potential interpretation provided in~\cite{Baybusinov2025}

\paragraph{Constraining the strengths of an entire layer} Having fully characterized the single-constraint case, we can now consider what happens when constraints are imposed on an entire layer rather than just on global mass. When constraints are imposed only on the strengths of one layer, denoted by the set $\{s_i\}$, the partition function factorizes into $M$ independent instances of the single-constraint model described above. Each instance corresponds to a specific node $i$ with its own local constraint $s_i$, obtained by substituting $\bar{W} \rightarrow s_i$ in the formulation of the global constraint. The overall partition function becomes the product of $M$ terms where the average mass per edge is substituted by the average strength per edge.

This factorization occurs because nodes in the constrained layer face independent optimization problems: node $i$ must distribute its mass $s_i$ across all links, but this decision does not restrict the choices available to other nodes since the receiving layer remains unconstrained. The equivalence between canonical and microcanonical ensembles established for the single-constraint case therefore carries over directly to the single-layer ensemble. This decomposition enables analytical tractability while preserving the essential statistical features of the single-constraint model.

However, this independence breaks down when constraints are imposed on both layers simultaneously. In the SOT model, each edge weight $W_{i\alpha}$ must satisfy constraints from both its source node and its target node (see Eq. \ref{eq:marginals}). These coupled constraints mean that node $i$'s decision to allocate mass to node $\alpha
$ directly affects the mass available for all other nodes connecting to $\alpha$. The partition function no longer factorizes, and we must develop more sophisticated mean-field techniques to access the system's macroscopic behavior.

\subsection{Full model}
\label{sec:full_model}
The full model with complete node-wise constraints requires introducing collective variables to separate global behavior from local fluctuations.
To access the macroscopic behavior of the partition function in Eq.~\eqref{eq:zsot}, we introduce collective (or \emph{relative}) variables that capture global properties while integrating out microscopic degrees of freedom.
This is accomplished systematically by inserting  \emph{resolutions of unity} that pin empirical averages via auxiliary Fourier fields. For instance, to constrain the total mass, we write:
\begin{equation}
	\label{eq:res_unity}
	1= \int d\omega\delta\left(\omega - \frac{1}{NM}\sum_{i,\alpha}W_{i\alpha}\right) = \int d\omega \,d\hat{\omega}\,e^{\imath\hat\omega\left(\omega - \frac{1}{NM}\sum_{i,\alpha}W_{i\alpha}\right)}
\end{equation}
This identity pins the empirical total mass to the scalar variable $\omega$, while the conjugate field $\hat\omega$ implements the Dirac's delta via Fourier transform.

The full model requires node-wise conservation at both layers. 
Beyond the total mass constraint in Eq.\eqref{eq:res_unity}, the model imposes $N$ constraints ensuring that each source node $i$ emits exactly $s_i$ units of mass and $M$ constraints ensuring that each target node $\alpha$ receives exactly $\sigma_\alpha$. Together, these three constraint types (one global and $N+M$ local) fully specify the ensemble of admissible mass distributions.

We start implementing the resolution of the unity related to the conservation of the total mass. Inserting Eq.(\ref{eq:res_unity}) into the left hand side of Eq.(\ref{eq:zsot}) and performing the integrals over each $W_{i\alpha}$ yields:
\begin{eqnarray}
	Z&=& \int d\omega d\hat{\omega} \mathcal{D}\lambda \mathcal{D}\mu \exp\left\{ \imath \sum_i\lambda_is_i +  \imath \sum_\alpha\mu_\alpha\sigma_\alpha  + \imath \hat\omega\omega \right.  \nonumber\\
	&\,& \left. - \sum_{i,\alpha}\ln(\beta C_{i\alpha} + \imath \lambda_i + \imath \mu_\alpha + \imath \frac{\hat{\omega}}{NM}) \right\} \label{eq:zsot_with_total_mass} \\
	&\,& \left(\prod_{i,\alpha} \Theta(\mathrm{Re}\{\beta C_{i\alpha} + \imath \lambda_i + \imath \mu_\alpha + \imath \frac{\hat{\omega}}{NM}\})\right) \nonumber
\end{eqnarray}
where the logarithmic terms originate from integrating $\int_0^\infty dW_{i\alpha} \, e^{-W_{i\alpha} r_{i\alpha}}$ with $r_{i\alpha} = \beta C_{i\alpha} + \imath\lambda_i + \imath\mu_\alpha$, and the product of Heaviside functions $\Theta(\cdot)$  ensures convergence by requiring $\lim_{W_{i\alpha} \to \infty} \exp\left( -W_{i\alpha} r_{i\alpha} \right) \to 0\quad$ which holds if and only if $\mathrm{Re}(r_{i\alpha}) > 0$. Integrating out $\hat\omega$ at this stage would simply re-impose the total-mass constraint:
\begin{equation*}
	\delta\left( \omega - \frac{1}{NM}\sum_{i,\alpha}\frac{1}{\beta C_{i\alpha} + \imath\lambda_i + \imath\mu_\alpha}\right)
\end{equation*}
consistent with the interpretation of $\omega$ as the intensive total mass variable.
However, we retain the integration over $\hat\omega$ for now to facilitate subsequent analysis.

The implementation of the node-level constraints requires decomposing the Lagrange multipliers into global and local components:
\begin{equation*}
	\lambda_i = l + x_i, \quad \mu_\alpha = m + y_\alpha, \quad l = \frac{1}{N} \sum_i \lambda_i, \quad m = \frac{1}{M} \sum_\alpha \mu_\alpha,
\end{equation*}
where $(l,m)$ are average fields (capturing the mean ``thermodynamic potentials'' of the two layers), while $\{x_i\}$ and $\{y_\alpha\}$ are zero-mean fluctuations satisfying $\sum_i x_i = 0$ and $\sum_\alpha y_\alpha = 0$ by construction and encoding node-level heterogeneities required to match the marginals $(s,\sigma)$.
This volume-preserving change of variables (unit Jacobian) enables a systematic saddle-point treatment by introducing resolutions of unity for the collective fields:
\begin{align}
	1 &= \int dl\, d\hat{l} \, e^{i\hat{l}(l - \frac{1}{N}\sum_i \lambda_i)} = \int dl \, d\hat{l} \, e^{-i\hat{l}\frac{1}{N}\sum_i x_i} \nonumber\\
	1 &= \int dm \, d\hat{m} \, e^{i\hat{m}(m - \frac{1}{M}\sum_\alpha \mu_\alpha)} = \int dm \, d\hat{m} \, e^{-i\hat{m}\frac{1}{M}\sum_\alpha y_\alpha} \nonumber
\end{align}
These identities make $(l,m)$ explicit integration variables and introduce conjugate fields $(\hat{l}, \hat{m})$ that enforce the averaging constraints. It follows that Eq.(\ref{eq:zsot_with_total_mass}) becomes:
\begin{eqnarray}
	Z&=& \int d\omega\, d\hat{\omega} \,\mathcal{D}x \,\mathcal{D}y \,dl\,d\hat{l} \,dm\,d\hat{m}\times \label{eq:before_log_expansion}\\ 
	&\,& \left(\prod_{i,\alpha} \Theta(\mathrm{Re}\{\beta C_{i\alpha} + \imath (l+m) + \imath x_i +\imath y_\alpha + \imath \frac{\hat\omega}{NM}\})\right) \times\nonumber\\
	&\,&\exp\left\{ \imath \bar{W}(l+m) + \imath \hat\omega\omega -\imath \frac{\hat{l}}{N}\sum_i x_i -\imath \frac{\hat{m}}{M}\sum_\alpha y_\alpha +\imath \sum_i x_i s_i\right. \nonumber \\
	&\,& \left. + \imath \sum_\alpha y_\alpha \sigma_\alpha - \sum_{i,\alpha}\ln(\beta C_{i\alpha} + \imath (l+m) + \imath x_i +\imath y_\alpha + \imath \frac{\hat\omega}{NM}) \right\}  \nonumber
\end{eqnarray}
This expression makes transparent the roles of global and local fields.
The combination $l+m$ acts as an effective \emph{global} potential shifting all logarithm arguments uniformly, thereby controlling overall mass loading, while the fluctuations $x_i$ and $y_\alpha$ implement the microscopic adjustments necessary to satisfy the node-wise constraints.

The real part of the argument of each logarithm must be positive to ensure convergence. Therefore we must require:
\begin{equation*}
	l+m\geq \max\{x_i +  y_\alpha +  \frac{\hat\omega}{NM} - \imath \beta C_{i\alpha}\}
\end{equation*}
This is automatically satisfied for $\beta>0$ and $C_{i\alpha}\ge 0$ (e.g., for costs drawn from a uniform distribution on $[0,1]$; a vanishing cost has measure zero and can be regularized if needed). Moreover, $x_i$ and $y_\alpha$ represent zero-mean fluctuations and, in a large system, are typically $O(1)$. In the limit $N, M \gg 1$, the contribution of $\hat{\omega}$ is suppressed by the factor $1/(NM)$ and is therefore negligible at leading order.

\begin{figure}[!t]
	\centering
	\includegraphics[width=1\linewidth]{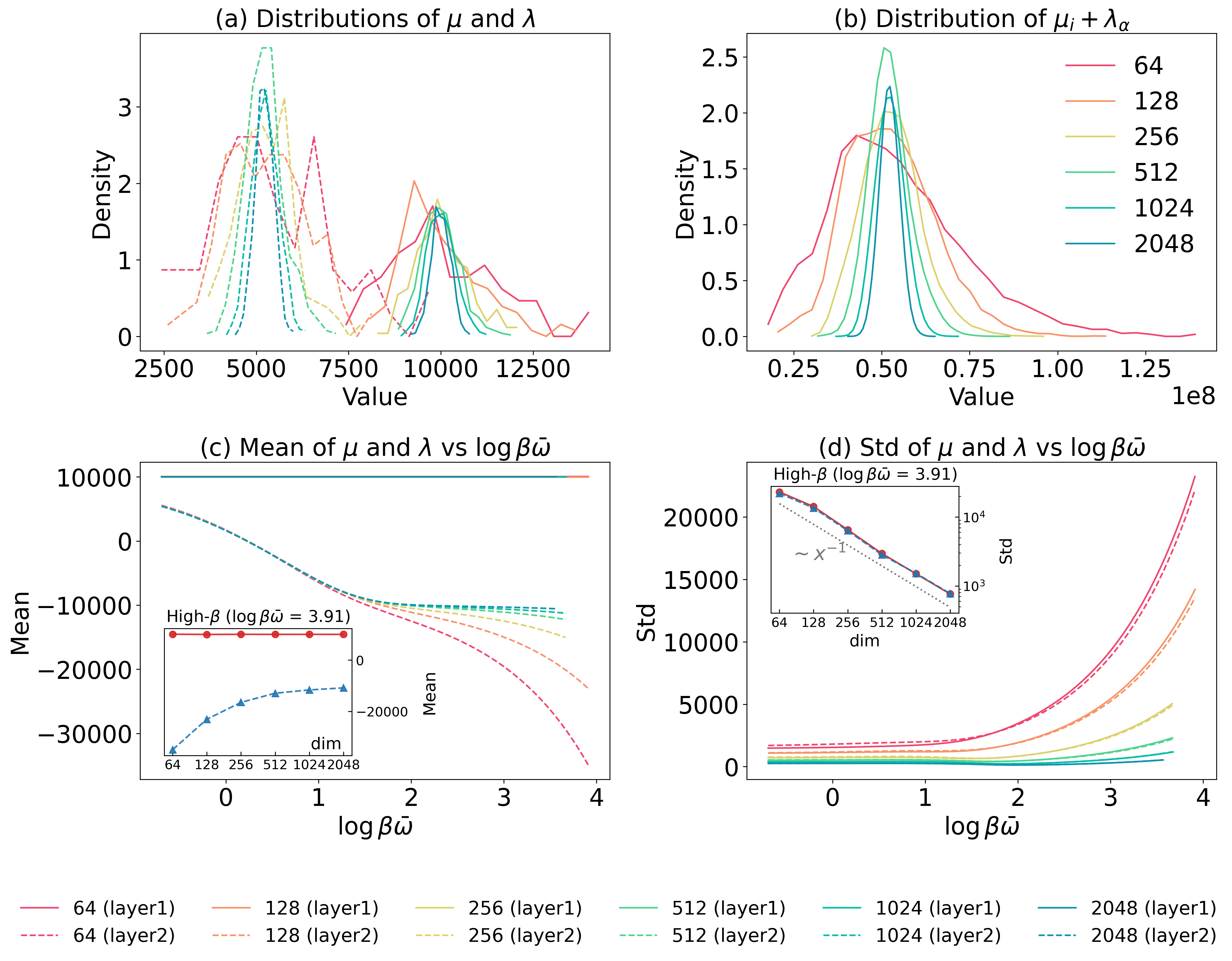}
	\caption{\textbf{Lagrange multipliers converge to Gaussian distributions in the thermodynamic limit and exhibit non-trivial dependence on $\beta$ and system size.} Numerical solutions of the SOT model with uniform cost distribution show that the distributions of individual multipliers $\lambda_i$ (solid lines, layer 1) and $\mu_\alpha$ (dashed lines, layer 2) approach Gaussian shape as system size increases (panel a), with the outer sum $\lambda_i + \mu_\alpha$ exhibiting even stronger convergence (panel b). (c) The mean multiplier values stabilize with increasing system size and become essentially constant at large $\beta$ (inset). (d) The multipliers' standard deviation grows with $\beta$ but decreases with system size, demonstrating that the thermodynamic limit and large-$\beta$ limit do not commute (inset shows the power-law saturation behavior $\text{Std}\sim \text{dim}^{-1}$). In the insets, colors distinguish between the two layers, not system sizes.}
	\label{fig:lagrange_multiplier}
\end{figure}

Before proceeding further, we examine numerical solutions of the SOT model to understand the behavior of local fluctuations. 
The partition function in its current form depends on $N+M$ fields variables whose distributions are not known \emph{a priori}. 
To guide our analytical treatment, we numerically solve the SOT model by finding the Lagrange multipliers $\{\lambda_i, \mu_\alpha\}$ that satisfy the saddle-point conditions for various system sizes and values of $\beta$. 
Figure~\ref{fig:lagrange_multiplier} presents the results across different regimes.

For large system sizes, the distributions of the multipliers $\lambda_i$ and $\mu_\alpha$ converge toward Gaussian shapes, as illustrated in panel (a). 
This behavior becomes more pronounced when considering the distribution of the outer sum $\lambda_i + \mu_\alpha$, shown in panel (b), which exhibits even stronger convergence to Gaussian form.

The mean and variance of these distributions reveal that the thermodynamic and large-$\beta$ limits do not commute.
Panels (c) and (d) present the corresponding means and variances as functions of $\log \beta\bar{\omega}$ for different system sizes. 
The mean approaches a stable value as system size increases, becoming essentially constant at very large $\beta$ (panel c). 
The inset of panel (c) shows a vertical slice at high $\beta$ (specifically at $\log \beta\bar{\omega} = 3.91$): it displays how the mean varies with system size at this fixed high coupling, confirming that the mean converges to a well-defined value in the thermodynamic limit.

The variance exhibits more intricate behavior: it increases with $\beta$ while decreasing with system size (panel d). 
The inset of panel (d) reveals the critical non-commutativity of limits through a vertical slice at high $\beta$: the variance follows a power-law decay with exponent $\sim -1$ as system size grows. 
This demonstrates that the order of limits matters crucially. 
If the large-$\beta$ limit is taken first ($\beta \to \infty$ before $N,M \to \infty$), the variance grows indefinitely without being compensated by system-size effects, and fluctuations remain large. 
Conversely, if the thermodynamic limit is taken first, the variance vanishes, and the Gaussian approximation becomes exact.
In what follows, we always take the thermodynamic limit first since we are interested in the finite $\beta$ regime. 

The Gaussian character of multiplier distributions observed in Figure~\ref{fig:lagrange_multiplier} justifies a systematic small-fluctuation expansion. 
The fluctuations $\{x_i, y_\alpha\}$ are also Gaussian with exponentially decaying tails. 
Combined with the power-law decrease of variance shown in panel (d), this ensures that $|x_i + y_\alpha|$ remains small relative to a background term for all edges simultaneously in the thermodynamic limit. 
We therefore expand the logarithm in Eq.\eqref{eq:before_log_expansion} by defining:
\begin{eqnarray*}
	B_{i\alpha}\;:=\;\beta C_{i\alpha}+\imath(l+m),
	\qquad
	\varepsilon_{i\alpha}\;:=\;\frac{\imath\,(x_i+y_\alpha+\hat\omega/NM)}{B_{i\alpha}}.
\end{eqnarray*}
so that:
\begin{eqnarray*}
	\ln\!\left(\beta C_{i\alpha}\;+\;\imath\bigl[(l+m)+x_i+y_\alpha+\hat\omega/NM \right)
	= \ln\bigl[B_{i\alpha}(1+\varepsilon_{i\alpha})\bigr] 
\end{eqnarray*}
For $\vert\varepsilon_{i\alpha}\vert\ll 1$ we use the Taylor expansion $ \ln(1+\varepsilon)=\varepsilon-\tfrac12\varepsilon^{2}+O(\varepsilon^{3})$ retaining only the linear term at leading order. 
The key assumption in our derivation is this small-fluctuation (small–$\varepsilon$) hypothesis, justified in the large-system, weak-heterogeneity regime (bounded $x_i,\,y_\alpha$ and finite $\beta C_{i\alpha}$). 
Outside this regime, higher-order terms should be retained, though we expect qualitative conclusions to remain unchanged.

The linear expansion around the smooth background $B_{i\alpha}$ cleanly separates global control, through $l+m$, from local adjustments, through $x_i$ and $y_\alpha$. The partition function becomes:
\begin{eqnarray}
	Z&=& \int d\omega\, d\hat{\omega} \,\mathcal{D}x \,\mathcal{D}y \,dl \,dm\,\Theta(\mathrm{Re}\{\beta C_{i\alpha} + \imath (l+m)\})\times  \nonumber\\
	&\,&\exp\left\{ \imath \bar{W}(l+m) + \imath \hat\omega\omega -  \sum_{i,\alpha}\ln\left(\beta C_{i\alpha} + \imath (l+m) \right) \right. \nonumber\\
	&\,&\left.  -\imath \frac{\hat\omega}{NM}\sum_{i,\alpha}\frac{1}{\beta C_{i\alpha} + \imath (l+m)}\right\} \Psi(l,m,s,\sigma,\beta C)
	\label{eq:Z_mf_small_epsilon}
\end{eqnarray}
where $\Psi(l,m,s,\sigma,\beta C)$ collects all the fluctuating terms involving Lagrange multipliers.
As discussed in \ref{app:sub-extensiveness}, $\Psi$ is sub-extensive in $(N\times M)$ and can therefore be neglected from the elements affecting the partition function in the thermodynamics limit. The partition function reduces to the one in Eq.\eqref{eq:7}:
\begin{eqnarray*}
	Z&\sim& \int d\lambda \,d\omega\,d\hat{\omega}\,\exp\left\{NM\left(\bar{\omega}\lambda - \frac{1}{NM}\sum_{i,\alpha}\ln(\beta C_{i\alpha} + \lambda)\right) \right.\\
	&\,&\left. + \imath \hat\omega\omega - \imath \frac{\hat\omega}{NM}\sum_{i,\alpha}\frac{1}{\beta C_{i\alpha} + \lambda}   \right\}\\
	&\sim&\int d\lambda\,\exp\left\{NM\left(\bar{\omega}\lambda - \frac{1}{NM}\sum_{i,\alpha}\ln(\beta C_{i\alpha} + \lambda)\right)\right\}
\end{eqnarray*}
This demonstrates that, in the mean-field framework, the partition function of the SOT model with all the strength constraints is dominated by the partition function of the model with only the global constraint, discussed in Section \ref{sec:one_constraint}. 
The full complexity of local fluctuations becomes irrelevant in the thermodynamic limit, and the essential physics is captured by the simpler global-constraint model.

\begin{figure}[t!]
	\centering
	\includegraphics[width=1\linewidth]{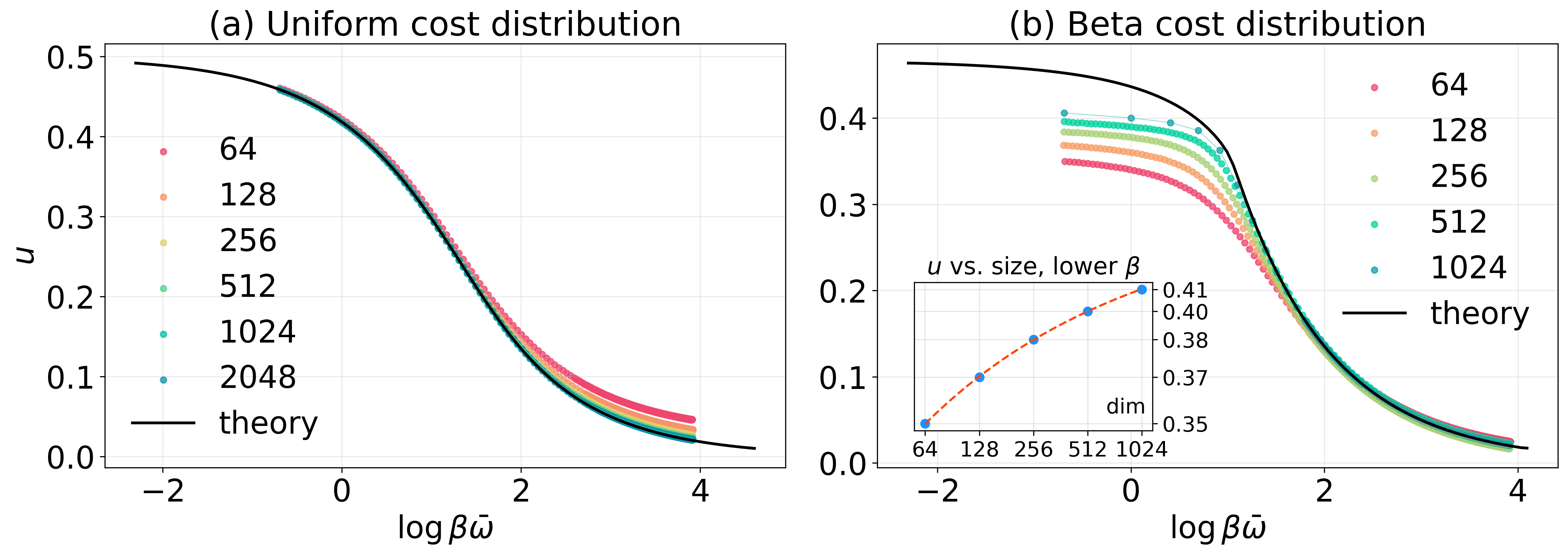}
	\caption{\textbf{Mean energy at different $\beta$:} Numerical results with (a) uniform and (b) Beta cost distributions. Colors encode the linear system size; the solid black line shows the theoretical predictions. The inset  of panel (b) shows how the mean energy varies as a function of the system size for the smallest available value of $\beta$; the red dashed line corresponds to the best fit using a saturating power-law $f(x\vert a,b,c) = a-b x^{-c}$, $a=0.45$, $b=0.35$, $c=0.31$.}
	\label{fig:energy}
\end{figure}
For costs drawn from a uniform distribution, the average energy converges to the prediction of Eq.\eqref{eq:energy}. 
Panel (a) of Figure~\ref{fig:energy} shows excellent agreement between numerical results and the theoretical prediction across all system sizes, in both the large and low $\beta$ regimes.

\subsection{Generalization to other homogeneous cost distributions}
The generalization to non-uniform but homogeneous cost distributions is possible using the ensemble equivalence, allowing the computation of the free energy through saddle point. 
For a general cost distribution $\rho(c)$, we can start from Eq.\eqref{eq:7} and \eqref{eq:glivenko_cantelli} since the partition function retains the form~\footnote{here we consider a complex rotation $\lambda\to\imath\lambda$.}:
\begin{equation}
	Z \sim \frac{1}{2\pi} \int d\lambda \exp\left\{N_\gamma\left[\bar{\omega}\lambda - \Phi(\lambda, \beta)\right]\right\}
\end{equation}
where:
\begin{equation}
	\Phi(\lambda, \beta) = \int \rho(c) \ln(\beta c + \lambda) dc
\end{equation}
The Saddle Point Approximation remains applicable since the large parameter $N_\gamma$ ensures the integral is dominated by the saddle point $\lambda^*$, determined by:
\begin{equation}
	0 = \frac{\partial}{\partial\lambda}[\bar{\omega}\lambda - \Phi(\lambda, \beta)]\Big\vert_{\lambda^*} \quad \Rightarrow \quad \bar{\omega} = \int \rho(c) \frac{1}{\beta c + \lambda^*} dc
\end{equation}
While this equation cannot be solved analytically for general $\rho(c)$, it can be obtained numerically. Crucially, the saddle point value always equals the susceptibility $\lambda^* = \chi(\beta, \bar{\omega})$ regardless of the choice of $\rho(c)$. 
This universal relation means that once $\lambda^*$ is determined numerically, the mean energy can be computed as $u = (1- \bar{\omega}\lambda^*)/\beta$ without further approximation.

Panel (b) of Figure~\ref{fig:energy} shows the comparison for costs drawn from a Beta distribution $\rho(c) \sim c^{\eta}$ with support $c \in [0,1]$. 
This behavior is representative of the broader class of cost laws with vanishing density at $c\to 0$, and might be therefore qualitatively distinct from the uniform case, which instead typifies distributions with finite $\rho(0)$ (see \cite{aldous2001zeta} for a short discussion).
In the simulations of the SOT model considered in~\cite{buffa2025}, the exponent $\eta$ depends on system size through $\eta = 1/\log_{10}(L)$, which explains the discrepancy observed in the dense regime where finite-size effects remain significant. 
As expected, the agreement becomes excellent in the large-$\beta$ regime while finite-size effects become apparent at lower $\beta$ values, but vanish as the size increases (see inset).

\section{Weight distribution in the ensemble and analysis in the large-$\beta$ regime}

The SOT model provides a framework to connect observable weight patterns to underlying optimization constraints. Notice that the cost structure driving these patterns may be unknown or uncertain. However, the weight distribution $\rho_W(w)$ and the cost distribution $\rho_C(c)$ are intrinsically linked: properties of the cost distribution manifest directly as observable features in the weight statistics. In this section we demonstrate that, for short-tails distributed costs, the weight distribution follows a power law $\rho_W(w) \sim w^{-2}$ in the large-$\beta$ regime.

We start from the relation between costs and weights. From Eq.\eqref{eq:zsot_with_total_mass}, where we introduced $r_{i\alpha} = \beta C_{i\alpha} + \imath\lambda_i + \imath\mu_\alpha + \imath\,\hat{\omega}/(NM)$, we derive an expression for the mean weight by taking the functional derivative of $\ln Z$ with respect to $r_{i\alpha}$. This yields $\langle w_{i\alpha} \rangle = 1/r_{i\alpha}$. After performing the imaginary rotation and neglecting the subdominant $\hat{\omega}/(NM)$ term in the thermodynamic limit, we obtain:
\begin{equation}
	\label{eq:w_avg}
	\langle w_{i\alpha} \rangle = \frac{1}{\beta C_{i\alpha} + \theta_{i\alpha}}, \quad \theta_{i\alpha} \equiv \lambda_i + \mu_\alpha
\end{equation}
an expression which is analogous to the one found in \cite{buffa2025}.
As established in the previous section, $\theta$ follows a Gaussian distribution whose fluctuations vanish as $N, M \to \infty$ (Figure~\ref{fig:lagrange_multiplier}, panel b), causing each weight $w_{i\alpha}$ to concentrate sharply around its ensemble average $\langle w_{i\alpha} \rangle$. This concentration allows us to replace the fluctuating weight with its mean value in the empirical distribution:
\begin{equation}
	\rho^{(\text{emp})}_W(w) = \frac{1}{N_\gamma} \sum_{i,\alpha} \langle \delta(w - w_{i\alpha})\rangle \approx \frac{1}{N_\gamma} \sum_{i,\alpha} \delta(w - \langle w_{i\alpha}\rangle)
\end{equation}
Therefore, the replacement $\langle \delta(w - w_{i\alpha})\rangle \to \delta(w - \langle w_{i\alpha}\rangle)$ increasingly accurate as $N, M \to \infty$. The empirical weight distribution thus becomes:
\begin{equation}
	\label{eq:empirical_rhoW}
	\rho^{(\text{emp})}_W(w) = \frac{1}{N_\gamma} \sum_{i,\alpha} \delta\left[ w - \frac{1}{\beta C_{i\alpha} + \theta_{i\alpha}} \right]
\end{equation}
At large $\beta$, the optimization selects a small subset of edges with minimal costs to carry most of the mass, while weights on the remaining unselected edges become vanishingly small. For these unselected edges, the weight is so small that it is plausibly independent of the specific cost value, what matters is only that the edge was not selected. Crucially, the number of selected edges scales linearly with system size $N+M$ (forming a spanning tree), while the number of unselected edges grows quadratically as $N\times M$. Since the empirical weight distribution is dominated by the much larger population of unselected edges, it is reasonable to assume that the Lagrange multipliers $\theta_{i\alpha}$ are uncorrelated with the corresponding costs $C_{i\alpha}$ in this regime. Under this assumption, the joint distribution factorizes $\rho(c, \theta) \approx \rho_C(c) \, \rho_{\theta}(\theta)$. The distribution $\rho_{\theta}(\theta)$ is the Gaussian established in Figure~\ref{fig:lagrange_multiplier} panel (b). This factorization allows us to compute the weight distribution $\rho_W(w)$ by convolving these independent distributions: 
\begin{eqnarray}
	\label{eq:rho_W_mix}
	\rho_W(w) &=& \int d\theta \, \rho_{\theta}(\theta) \int dc \, \rho_{C} (c) \, \delta\left(w - \frac{1}{\beta c + \theta}\right) \nonumber\\
	&=& \frac{1}{w^2}\int d\theta \, \rho_{\theta}(\theta) \int dc \, \rho_{C} (c) \, \delta\left(\theta - \frac{1}{w} + \beta c\right) \nonumber\\
	&\approx& \frac{1}{w^2}\int dc \, \rho_{C} (c) \, \delta\left(\bar\theta - \frac{1}{w} + \beta c\right)
\end{eqnarray}
In the last step, the $\theta$ distribution is approximated by a Dirac delta $\delta(\theta-\bar\theta)$, where $\bar\theta$ is the Gaussian mean. 
Since the Dirac delta constrains $c$, to satisfy $c = (1-\bar\theta w)/\beta w$, we must ensure this expression remains within the prescribed integration limits.  Given that the cost matrix entries are lower bounded ($c\geq 0)$, the condition $w \leq 1/\bar\theta$ must be satisfied, yielding a power-law weight distribution:
\begin{equation}
	\label{eq:change_w_to_c}
	\rho_W(w) =
	\begin{cases}
		\displaystyle \rho_{C} \left( \frac{1 - \bar\theta w}{\beta w} \right) \times \frac{1}{\beta w^2} & \qquad \displaystyle w \leq  \frac{1}{\bar\theta } \\[6pt]
		0 & \qquad \text{otherwise}
	\end{cases}
\end{equation}

\begin{figure}[!t]
	\centering
	\includegraphics[width=1\linewidth]{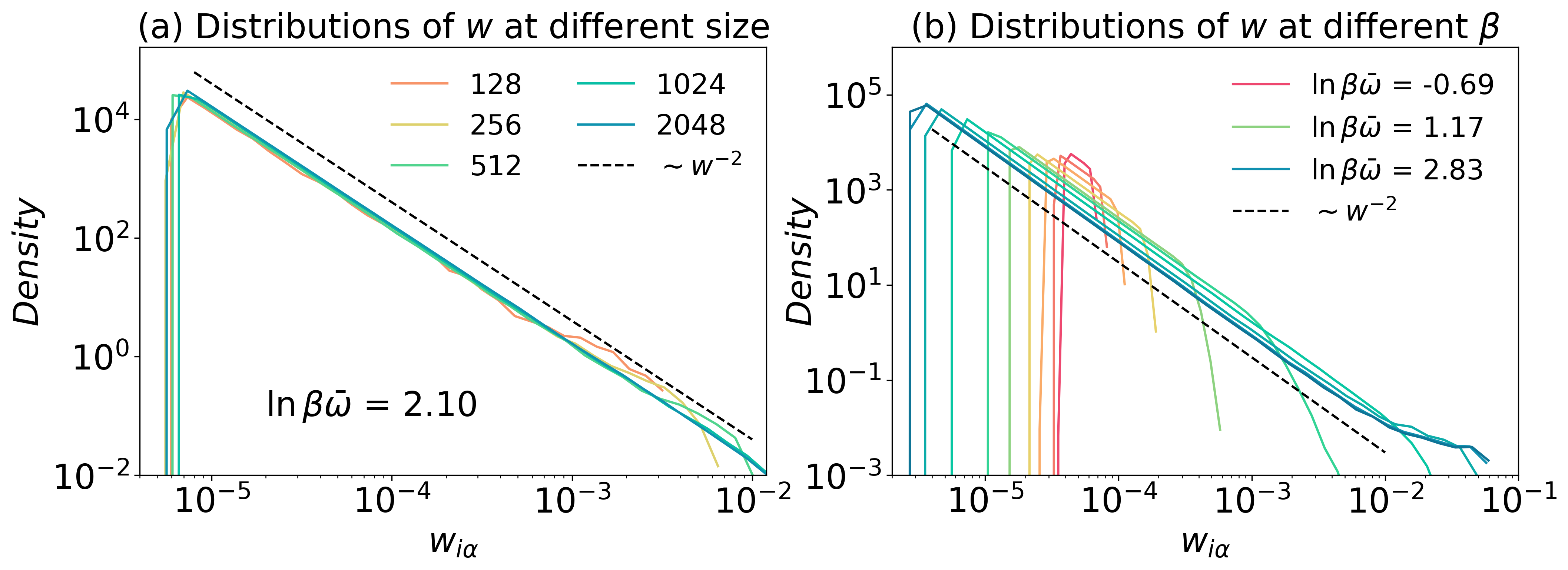}
	\caption{\textbf{Distribution of the weights for the uniform cost:} Panel (a) shows the distribution of $w$ from the numerical solution of the SubOT model at different system sizes.
		Panel (b) shows the distribution of $w$ varying the parameter $\beta$, thus moving from the dense regime at low $\beta$ to the sparse regime at large $\beta$.}
	\label{fig:distribution_weights}
\end{figure}
To test Eq.\eqref{eq:change_w_to_c} we generated several weight configurations by numerically solving the SOT model in \cite{buffa2025} with uniformly distributed costs on $[0,1]$ at various values of $\beta$. These weights make up the empirical distribution defined in Eq.\eqref{eq:empirical_rhoW} that we compare with the theoretical prediction from Eq.\eqref{eq:change_w_to_c}. Figure~\ref{fig:distribution_weights} shows excellent agreement between the empirical distributions (solid lines) and analytical predictions (dashed lines) across different system sizes (panel a) and values of $\beta$ (panel b). The distributions overlap within the interval $w \in [1/(\bar\theta + \beta), 1/\bar\theta]$.

We further observe that as $\ln \beta$ increases, we have the variance of $\theta$ to vanish, as shown in panel (b) of Fig.~\ref{fig:lagrange_multiplier}. When $w < 1/(t + \beta)$, the solution reduces to the power-law form $\rho_W(w) = 1/(\beta w^2)$.  This prediction is confirmed in panel (a) of Fig.~\ref{fig:distribution_weights}, where the weight distributions follow the expected power-law decay (with exponent -2) for each system size. They also exhibit a sharp lower cutoff at $w \geq 1/\beta$. At large $\beta$, this is justified by $\bar\theta\ll \beta$, allowing the approximation  $\bar\theta \approx 0$ to hold.

At lower $\beta$, deviations from ideal power-law scaling emerge as the mean multiplier becomes comparable to $\beta$. When $\beta$ decreases, $\bar\theta$ becomes non-negligible relative to $\beta$, shifting the lower bound from $1/\beta$ to $1/(\bar\theta + \beta)$. This shift is evident in panel (b), where minimum weights in the average matrix fall below $1/\beta$. Deviations from the ideal $1/(\beta w^2)$ scaling become increasingly pronounced, reflecting growing correlation between $\theta$ and the cost matrix $C$. Nevertheless, Eq.\eqref{eq:change_w_to_c} continues to accurately describe a portion of the weight distribution even at low $\beta$, indicating that decorrelation holds over a shrinking subset of edges. This behavior is consistent with our hypothesis: as $\beta$ increases, a larger fraction of edges becomes independent of cost values, broadening the regime where the analytical prediction remains valid.

\section{Conclusion}
\label{sec:conclusion}

In this work, we investigated the Sub-Optimal Transport problem from a statistical mechanics perspective, developing an analytical framework to understand the macroscopic behavior of the SOT model. 
Our primary goal was to move beyond purely numerical observations by deriving a tractable theoretical description capable of explaining the dense-to-sparse structural transition that emerges as a function of the tuning parameter $\beta$.

We began by studying a simplified model through a relaxation of the constraints, replacing the original layered conservation conditions with a single global constraint on the total transported mass. This reduction enabled an explicit mean-field solution for the case in which the cost matrix is independently and uniformly distributed. Despite its simplicity, this reduced model proved remarkably informative: it allowed us to fully characterize several key phenomena previously reported only in numerical studies \cite{buffa2025}. In particular, we analytically explained the logarithmic scaling in $\beta$, the smooth and non-critical nature of the dense-to-sparse crossover, and the equivalence between canonical and microcanonical ensembles in the thermodynamic limit. These results clarify that the transition does not correspond to a conventional phase transition but rather to a continuous structural reorganization driven by competing entropic and cost-based forces.

Building on this foundation, we then tackled the full constrained model, reinstating conservation conditions on both layers. Using a saddle-point approximation under the same assumptions on the cost matrix, we derived self-consistent equations governing the Lagrange multipliers. To close the system, we leveraged numerical evidence to justify a Gaussian approximation for the distribution of the multipliers at large system sizes. This approach allowed us to preserve analytical tractability while retaining the essential structural features of the original model. We further generalized our results to broader families of cost distributions, demonstrating that our framework applies beyond the uniform case, provided that correlations in the cost matrix remain negligible.

A key conceptual departure from much of the existing literature is that we do not focus on the strict optimal-transport limit ($\beta \to \infty$). Instead, we emphasize the physically and practically relevant regime of finite $\beta$, where solutions remain sub-optimal, and the interplay between randomness and optimization is most pronounced. Within this regime, we show that the limits ($N \to \infty$) and ($\beta \to \infty$) do not commute, revealing a subtle but important structural property of the model and underscoring the necessity of treating large-scale and strong-optimization limits with care.

Finally, we established an explicit relationship between the cost distribution and the induced distribution of transport weights, providing a direct link between microscopic cost statistics and macroscopic transport patterns. This result opens the door to inverse problems, where empirical transport networks could be used to infer underlying cost landscapes.

Overall, our findings demonstrate that statistical mechanics provides a powerful and unifying framework for studying sub-optimal transport, offering analytical insight into regimes that are difficult to access by optimization theory alone. 
Beyond the SOT model, we expect our approach to be broadly applicable to other systems in which entropy, cost, and structural constraints compete, including infrastructure networks, biological transport systems, and economic flow models.

\section*{Acknowledgemens}
D.M., A.P. and R.P. acknowledge the financial support under the National Recovery and Resilience Plan (NRRP), Mission 4, Component 2, Investment 1.1, Call for tender No. 104 published on 2.2.2022 by the Italian Ministry of University and Research (MUR), funded by the European Union – NextGenerationEU– Project Title  "WECARE – WEaving Complexity And the gReen Economy" – CUP 20223W2JKJ by the Italian Ministry of University and Research (MUR). 

\bibliographystyle{elsarticle-num} 
\bibliography{biblio.bib}

\newpage
\appendix
\section{Derivation of the mean-field partition function in Eq.(\ref{eq:Z_mean_field})}
\label{app:derivation_Z_mean_field}

This appendix details the derivation of the closed-form partition function in Eq.\eqref{eq:Z_mean_field} from the contour integration result in Eq.\eqref{eq:4c}. The starting point is the partition function expressed via residue calculus:
\begin{equation}
	\label{eq:A1}
	Z = \frac{1}{\beta^{N_\gamma-1}} \sum_{\gamma} (-1)^\gamma e^{-\beta \bar{W} C_\gamma} \prod_{\gamma' \neq \gamma} \frac{1}{|C_{\gamma'} - C_\gamma|}
\end{equation}

The key step in obtaining a tractable expression is the mean-field approximation for the ordered cost variables. When the costs $C_\gamma$ are independently drawn from a uniform distribution on $[0,1]$ and arranged in increasing order such that $C_\gamma < C_{\gamma+1}$ for $\gamma \in [1, \ldots, N_\gamma]$, the ordered variables become self-averaging in the large-$N_\gamma$ limit. Under this assumption, each ordered cost can be replaced by its expected value, which corresponds to the $\gamma$-th order statistic of $N_\gamma$ independent uniform samples:
\begin{equation}
	C_\gamma \approx \mathbb{E}[C_\gamma] = \frac{\gamma}{N_\gamma + 1}
\end{equation}

This approximation allows explicit evaluation of the product over cost differences appearing in Eq.\eqref{eq:A1}. Considering $|C_{\gamma'} - C_\gamma| = |\gamma' - \gamma|/(N_\gamma + 1)$ yields:
\begin{equation}
	\prod_{\gamma' \neq \gamma} \frac{1}{|C_{\gamma'} - C_\gamma|} = \frac{(N_\gamma + 1)^{N_\gamma - 1}}{\prod_{\gamma' \neq \gamma} |\gamma' - \gamma|}
\end{equation}

Without loss of generality we can factorize the product $\prod_{\gamma' \neq \gamma} |\gamma' - \gamma|$ into contributions where the difference between $\gamma$ and $\gamma'$ is positive or negative. Say $\gamma-1$ is the last edge for which $\gamma' < \gamma$. Those terms are $(\gamma - 1), (\gamma - 2), \ldots, 1$, which multiply to $(\gamma - 1)!$. For $\gamma' > \gamma$, the terms are $1, 2, \ldots, (N_\gamma - \gamma)$, yielding $(N_\gamma - \gamma)!$. Combining these results:
\begin{equation}
	\prod_{\gamma' \neq \gamma} |\gamma' - \gamma| = (\gamma - 1)!\,\, (N_\gamma - \gamma)!
\end{equation}

The factorial expression can be rewritten in terms of the binomial coefficient by noting that:
\begin{equation}
	\frac{1}{(\gamma - 1)!(N_\gamma - \gamma)!} = \frac{\gamma}{\gamma!(N_\gamma - \gamma)!} = \frac{\gamma}{N_\gamma!} \binom{N_\gamma}{\gamma}
\end{equation}

The exponential factor in Eq.\eqref{eq:A1} also simplifies under the mean-field approximation. Substituting $C_\gamma = \gamma/(N_\gamma + 1)$ and introducing the intensive variable $\bar{\omega} = \bar{W}/N_\gamma$:
\begin{equation}
	e^{-\beta \bar{W} C_\gamma} = \exp\left(-\beta \bar{W} \frac{\gamma}{N_\gamma + 1}\right) \approx e^{-\beta \bar{\omega} \gamma}
\end{equation}
where the approximation becomes exact in the thermodynamic limit $N_\gamma \to \infty$. Assembling these results and noting that in the large-$N_\gamma$ limit the prefactors combine to give $1/\beta^{N_\gamma}$, the partition function reduces to:
\begin{equation}
	Z \simeq \frac{1}{\beta^{N_\gamma}} \sum_{\gamma=0}^{N_\gamma} (-1)^\gamma e^{-\beta \bar{\omega} \gamma} \binom{N_\gamma}{\gamma}
\end{equation}
This sum has a closed-form expression through the binomial theorem. Recognizing the expansion of $(1 - x)^n = \sum_{k=0}^{n} \binom{n}{k} (-1)^k x^k$ with $x = e^{-\beta\bar{\omega}}$ and $n = N_\gamma$:
\begin{equation}
	\sum_{\gamma=0}^{N_\gamma} \binom{N_\gamma}{\gamma} (-1)^\gamma e^{-\beta\bar{\omega}\gamma} = \left(1 - e^{-\beta\bar{\omega}}\right)^{N_\gamma}
\end{equation}
The partition function therefore, takes the compact form:
\begin{equation}
	Z \simeq \frac{1}{\beta^{N_\gamma}} \left(1 - e^{-\beta\bar{\omega}}\right)^{N_\gamma}
\end{equation}
which is precisely Eq.\eqref{eq:Z_mean_field}.

\section{Sub-extensivity of $\Psi$}
\label{app:sub-extensiveness}
Integrating over $(\hat{l},\hat{m})$, there are 2 deltas, and we have the integral over the $\{x_i,y_\alpha\}$ that can be written as
\begin{eqnarray*}
	\Psi(l,m,s,\sigma,\beta C)&\simeq&\int \mathcal{D}x \,\mathcal{D}y\,\delta\left(\frac{1}{N}\sum_ix_i\right)\,\delta\left(\frac{1}{M}\sum_\alpha y_\alpha\right) \times \\
	&\,&\exp\left\{ - \imath\sum_i x_i \left(\sum_{\alpha}w_{i\alpha} -s_i\right)- \imath\sum_\alpha y_\alpha\left(\sum_{i}w_{i\alpha} - \sigma_\alpha\right)\right\}\\
	&\sim& \int d\hat{l}d\hat{m} \left(\prod_i e^{-\imath x_i(\sum_\alpha w_{i\alpha} - s_i - \frac{\hat{l}}{N})}\right)\left(\prod_\alpha e^{-\imath y_\alpha(\sum_i w_{i\alpha} - \sigma_\alpha - \frac{\hat{m}}{M})}\right)\\
	&\sim& \left(\prod_i \delta(\sum_\alpha w_{i\alpha} - s_i)\right)\left(\prod_\alpha \delta(\sum_i w_{i\alpha} - \sigma_\alpha )\right)
\end{eqnarray*}
where $w_{i\alpha}=\frac{1}{\beta C_{i\alpha} + \imath (l+m)}$, and there are $M+N$ components in $\Psi$. 
Fluctuations are of order $1/N$ and $1/M$.

\end{document}